\documentclass[pra,aps,twocolumn,showpacs, floatfix]{revtex4}
\usepackage{amssymb}
\usepackage{graphicx}
\usepackage{dcolumn}
\usepackage{bm,amsmath,verbatim}

\def\be{\begin{equation}}
\def\ee{\end{equation}}
\def\bea{\begin{eqnarray}}
\def\eea{\end{eqnarray}}
\def\bse{\begin{subequations}}
\def\ese{\end{subequations}}

\def\be{\begin{eqnarray}}
\def\ee{\end{eqnarray}}

\begin{document}

\title{Bose-Einstein Condensates in Spin-Orbit Coupled Optical Lattices:
Flat Bands and Superfluidity}
\author{Yongping Zhang$^{1}$}
\author{Chuanwei Zhang$^{1,2}$}
\thanks{Corresponding Author, Email: chuanwei.zhang@utdallas.edu}
\affiliation{$^{1}$Department of Physics and Astronomy, Washington State University,
Pullman, WA, 99164 USA \\
$^{2}$Department of Physics, the University of Texas at Dallas, Richardson,
TX, 75080 USA
}

\begin{abstract}
Recently spin-orbit (SO) coupled superfluids in free space or harmonic traps
have been extensively studied, motivated by the recent experimental
realization of SO coupling for Bose-Einstein condensates (BEC). However, the
rich physics of SO coupled BEC in optical lattices has been largely
unexplored. In this paper, we show that in suitable parameter region the
lowest Bloch state forms an isolated flat band in a one dimensional (1D) SO
coupled optical lattice, which thus provides an experimentally feasible
platform for exploring the recently celebrated topological flat band physics
in lattice systems. We show that the flat band is preserved even with the
mean field interaction in BEC. We investigate the superfluidity of the BEC
in SO coupled lattices through dynamical and Landau stability analysis, and
show that the BEC is stable on the whole flat band.
\end{abstract}

\maketitle

\section{introduction}

Flat bands possess macroscopic level degeneracy because of their flat energy
dispersion. They play a crucial role in important physical phenomena such as
fractional quantum Hall effects where a large magnetic field applied on a
two dimensional electron gas induces flat Landau levels \cite{Landau}. The
flat band physics is also greatly enriched recently by studying various
lattice models where flat bands can be generated through geometrical
frustration of hopping \cite{Harris,Ohgushi} (e.g. Kagome lattice), the
destructive interference between nearest-neighbor and higher-order
tunnelings (such as next-nearest-neighbor) \cite{Weaire,Straley,Tasaki}, or
the \textit{p}-orbital physics \cite{WuCongjun}. In particular, isolated
flat bands in lattices with non-trivial topological properties have
attracted much attention in condensed matter physics for their applications
in engineering fractional topological quantum insulator \cite%
{Neupert,Sheng,Sun,Tang,Wang,Xiao,YFW} without Landau levels.

However, most previous lattice models for generating flat bands involve
either high orbital bands or high order tunnelings, which are generally very
challenging in experiments. In this paper, we propose an experimentally
feasible route for generating isolated flat bands using cold atoms in SO
coupled weak optical lattices. Our work is motivated by the recent
experimental realization of SO coupling for BEC \cite{Lin}, which opens a
completely new avenue for exploring SO coupled superfluids \cite{Dalibard}.
In particular, SO coupled BEC and degenerated fermi gases in free space and
harmonic traps have been extensively investigated recently \cite%
{Zhai,Ho,Wucongjun2,Yip,Xu,Kawakami,Xiaoqiang,Zhou,Radic,Sinha,Hu,Yun,Ramachandhran,Ozawa,Yongping1,Yongping2,Zhu,Bijl,Chen,Larson1,DFG1,DFG2,DFG3}%
. However, ultra-cold atoms in SO coupled optical lattices have been largely
unexplored \cite{Larson2}. We show that the combination of SO coupling,
Zeeman field and optical lattice potential can yield isolated flat bands
where topological properties may originate from the SO coupling \cite%
{Roman,Oreg}. In regular optical lattices, the minimum of the lowest Bloch
band locates at the center of the first Brillouin zone (BZ), while the
maximum at the edge \cite{Aschcoft}. In SO coupled optical lattices, the
minimum may locate at the edge and the peak at the center. The height of the
central peak can be reduced with increasing Zeeman field, leading to
decreasing band width and flat bands in certain parameter region. We note
that such flat band dispersion has been observed very recently in
experiments in SO coupled optical lattices using $^{6}$Li Fermi atoms \cite%
{Martin}.

We first investigate a single atom in a 1D SO coupled weak optical lattice
to illustrate the mechanism for generating isolated flat bands. The
atom-atom interaction in BEC is then taken into account using the mean-field
Gross-Pitaevskii (G-P) equation \cite{Pethick}. The nonlinear interaction
reduces the band flatness, but does not fully destroy the isolated flat
bands. The combination of nonlinear interaction and flat bands may lead to
rich and interesting physics. In particular, the instability of the
nonlinear Bloch waves is very important because it directly relates to the
breakdown of superfluidity of the BEC \cite%
{BiaoWu1,BiaoWu2,Smerzi,Machholm,Burger,Fallani,Morsch,Hooley,Dutton,Matt,Modugno}%
. In SO coupled optical lattices, the non-zero momentum of the energy
minimum of the lowest Bloch band and the existence of flat bands make their
stability very different from regular optical lattices. For instance, the
nonlinear Bloch waves can be stable in the whole BZ in the flat band region.

The rest of the manuscript is organized as follows. In Sec. II, we present
the flat band structure in SO coupled optical lattices. In Sec. III, we
discuss the effects of mean-field interactions and analyzes the stability of
the BEC in SO coupled optical lattices. Sec. IV is the conclusion.

\section{Flat bands in SO coupled optical lattices}

\begin{figure}[t]
\includegraphics[width=0.6\linewidth]{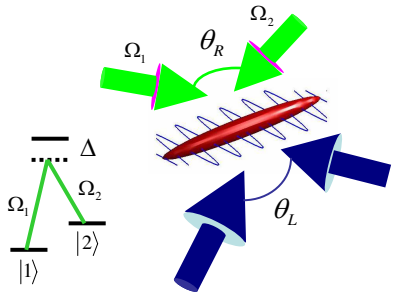}
\caption{(Color Online) Laser setup for implementing 1D SO coupled optical
lattices. $\Omega _{1}$, and $\Omega _{2}$ are the Rabi frequencies of the
Raman lasers for generating SO coupling. The other two laser beams generate
the optical lattice. }
\label{setup}
\end{figure}

We consider a BEC confined in a 1D optical lattice potential $V_{0}\sin
^{2}(k_{L}x)$ along the \textit{x} direction with $V_{0}$ as the lattice
depth. In experiments, the lattice potential can be created by a standing
wave formed by two lasers propagating along different directions \cite%
{Morsch} (see Fig. \ref{setup}). The effective wavevector of the lattice $%
k_{L}=2\pi \sin (\theta _{L}/2)/\lambda _{L}$, where $\lambda _{L}$ is the
wavelength of the lasers and $\theta _{L}$ is the angle between two lasers.
The SO coupling for BEC has been realized in experiments using two
counter-propagating Raman lasers \cite{Lin}, yielding the single particle
Hamiltonian
\begin{equation}
H_{0}=\frac{p^{2}}{2m}+\gamma p\sigma _{z}+\Omega \sigma _{x},
\end{equation}%
where $p$ is the atom momentum along the \textit{x} direction, and $\sigma $
is the Pauli matrix. The SO coupling strength $\gamma =\hbar k_{R}/m$ with $%
k_{R}=2\pi \sin (\theta _{R}/2)/\lambda _{R}$, $\lambda _{R}$ is the
wavelength of the Raman lasers, and $\theta _{R}$ is the angle between Raman
beams. $\Omega $ is the Rabi frequency and acts as a Zeeman field. The units
of the energy and length are chosen as the recoil energy $2E_{L}=\hbar
^{2}k_{L}^{2}/m$, and $1/k_{L}$ respectively for the numerical calculation.
Under these units, the single-particle Hamiltonian is dimensionless with $%
\gamma =k_{R}/k_{L}=\sin (\theta _{R}/2)\lambda _{L}/\sin (\theta
_{L}/2)\lambda _{R}$ and the optical lattice potential $V_{0}\sin ^{2}(x)$.

\begin{figure}[t]
\includegraphics[width=1\linewidth]{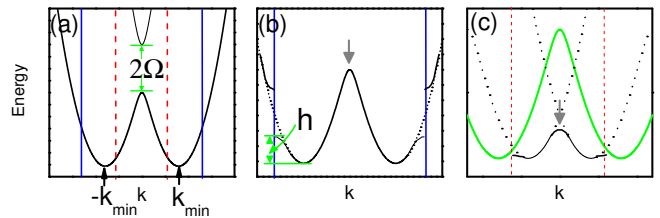}
\caption{(Color Online) Illustration of the formation of isolated flat
bands. (a) Energy dispersion $\protect\mu _{\pm }(k)$ of the single-particle
Hamiltonian $H_{0}$. The vertical solid (or dashed) lines correspond to the
edges of the first BZ. (b) Lowest band for $\protect\gamma <1$. The gray
arrow indicates the suppression of the central peak with increasing $\Omega $%
. (c) The formation of the lowest band for $\protect\gamma \geq 1$. The
solid (green) line is $\protect\mu _{-}(k)$, while the dotted lines are $%
\protect\mu _{-}(k\pm 2)$. }
\label{idea}
\end{figure}

Without optical lattice potentials, the single particle Hamiltonian $H_{0}$
has two SO energy bands $\mu _{\pm }(k)$ (shown in Fig. \ref{idea}a) due to
the lift of the spin degeneracy by the SO and Zeeman field. A gap $2\Omega $
between these two bands is opened at $k=0$ by the Zeeman field $\Omega $. In
the lower band, there are two energy minima at $k_{min}=\pm \sqrt{\gamma
^{2}-\Omega ^{2}/\gamma ^{2}}$ and one peak at $k=0$. With increasing $%
\Omega $, the distance between two $k_{min}$ shrinks, and the height of the
central peak decreases. At a critical value $\Omega _{c}=\gamma ^{2}$ and
beyond, two $k_{min}$ merge to one point at $k_{min}=0$, and the central
peak vanishes.

In the presence of periodic lattice potentials (\textit{i.e.}, consider the
Hamiltonian $H_{0}+V_{0}\sin ^{2}(x)$), the eigenenergies of the
single-particle Hamiltonian form the Bloch energy bands \cite{Aschcoft}. To
generate an isolated flat band, it is necessary to reduce the energy at both
the edge and the center of the first BZ with respect to the band minimum,
which can be realized through a combination of SO coupling, Zeeman field and
lattice potential. Specifically, the periodic lattice potential can open an
energy gap at the edge of the first BZ, which lowers the energy difference
between the edge and the band minimum (denoted as $h$). When the original
band minimum is close to the edge, the band edge becomes the band minimum (%
\textit{i.e.}, $h=0$). On the other hand, the height of the central peak
decreases with increasing $\Omega $. The band width should be determined by
the larger value of $h$ and the central peak height.

The flat band generation mechanism is slightly different in two different
regions: $\gamma <1$ and $\gamma \geq 1$. For $\gamma <1$, $k_{min}=\gamma $
at $\Omega =0$ is within the first BZ (Fig. \ref{idea}b). If $\gamma $ is
close to 1, $h$ should be zero and the band width is determined by the
central peak height, which can be greatly reduced with increasing $\Omega $.
Therefore the lowest band could be very flat for certain parameter region.
However, if $\gamma $ is much smaller than 1, $h$ becomes a large value, and
the width of the lowest band cannot be squeezed to the flat region. For $%
\gamma \geq 1$, $k_{min}$ of $H_{0}$ lays outside of the first BZ. In this
case, the lowest band is formed through folding the energy spectrum into the
first BZ (i.e., shift the energy band of $H_{0}$ by a lattice vector $k\pm 2$%
, see Fig. \ref{idea}c). The band minima now locate at $k_{min}\mp 2$, and
the physics is similar as that in $\gamma <1$. However, there is one major
difference between $\gamma \geq 1$ and $\gamma <1$. For $\gamma \geq 1$, the
minimum of the lowest band first shift towards the edge of the first BZ when
$\Omega $ increases from 0. Therefore at certain range of $\Omega $, the
band minimum always stays at the band edge and the flat band can be realized
by suppressing the central peak with increasing $\Omega $. We emphasize that
the resulting flat band is the ground state of the SO coupled lattice, which
further enhances its experimental feasibility because atoms are usually
adiabatically loaded to the lowest band in experiments \cite{Morsch}. Such
SO mechanism for flat bands is very different from previous schemes in
literature using high order tunneling or high orbital physics.

\begin{figure}[t]
\includegraphics[width=1\linewidth]{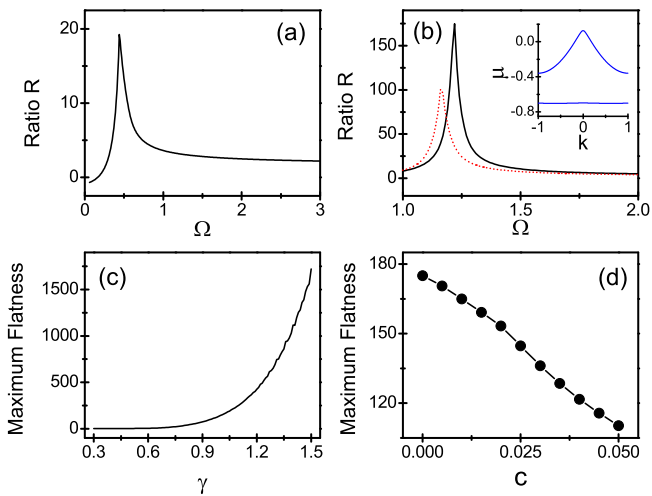}
\caption{(Color Online) Flatness (ratio $R$) of the lowest Bloch band. $%
V_{0}=1$. (a) $\protect\gamma =0.74$ and $c=0$. (b) $\protect\gamma =1.05$.
Solid line: $c=0$; Dashed line: $c=0.05$. The inset is a typical example of
flat band in the nonlinear Bloch spectrum with $\Omega =1.15$, $c=0.05$. (c)
Plot of the dependence of the maximum flatness on $\protect\gamma $ with $%
c=0 $. (d) The dependence of the maximum flatness on $c$ with $\protect%
\gamma =1.05$. }
\label{flatness}
\end{figure}

The above intuitive physical picture agrees well with the numerical results.
Using the Bloch theorem, the Bloch waves can be written as $\Psi(x,t) =\Phi
(x) \exp (-i\mu \left( k\right) t+ikx)$, where $\Phi (x)$ is the periodic
part of the Bloch wavefunction, and $\mu \left( k\right) $ is the
eigenenergy, which can be calculated using the standard central equation. We
measure the flatness of the lowest Bloch band by the ratio $R$ of the gap
between the lowest and first excited bands to the width of the lowest band.
In Fig. \ref{flatness}, we plot the flatness $R$ with respect to $\Omega $
for two different $\gamma $. In the calculation, we use $\lambda _{R}=804.1$
nm and $\lambda _{L}=840$ nm which are typical for $^{87}$Rb atoms in
experiments \cite{Lin}. The optical lattice potential is weak $V_{0}=2E_{L}$
to make sure the flat band does not come from the high lattice potential.
For simplicity we choose $\theta _{L}=\pi $, and consider two different $%
\theta _{R}$: $\theta _{R}=\pi $ corresponds to $\gamma =0.74$ (in Fig. \ref%
{flatness}a), and $\theta _{R}=\pi /2$ corresponds to $\gamma =1.05$ (in
Fig. \ref{flatness}b). We see the maximum flatness can reach nearly 20/1 for
$\gamma =0.74$ and 170/1 for $\gamma =1.05$. The suppression of the flatness
for $\gamma =0.74<1 $ agrees with our intuitive physical picture: the band
minimum for $\gamma =0.74$ is a little bit far from the edge of the first
BZ, therefore the lowest band cannot be squeezed to exactly flat. In
experiments, $\gamma $ can be varied using laser setups with different $%
\theta _{L}$ and $\theta _{R}$ or through a fast modulation of the laser
intensities of the Raman lasers \cite{Yongping2}. The dependence of the
maximum flatness on the SO coupling $\gamma $ is plotted in Fig. \ref%
{flatness}c. With increasing SO coupling, the lowest band becomes more flat.

\section{Stability of BEC in SO coupled optical lattices}

So far the study has been limited to the linear case, i.e., a single atom,
while the interactions between atoms in BEC may play a major role on the
dynamics of BEC. In the presence of a weak lattice potential, the mean field
theory still applies and the dynamics of BEC in SO coupled optical lattices
can be described by the non-linear G-P equation
\begin{equation}
i\frac{\partial \Psi }{\partial t}=H_{0}\Psi +V_{0}\sin ^{2}(x)\Psi +c(|\Psi
_{1}|^{2}+|\Psi _{2}|^{2})\Psi ,
\end{equation}%
where $\Psi =(\Psi _{1},\Psi _{2})^{T}$ is the two component wavefunction of
the BEC. The unit of time is $m/\hbar k_{L}^{2}$ and the wavefunction is
normalized through $\int dx(|\Psi _{1}|^{2}+|\Psi _{2}|^{2})=1$ in one unit
cell. The dimensionless interaction coefficient $c=\hbar \sqrt{\omega
_{y}\omega _{z}}k_{L}aN/E_{L}$, where $N$ is the atom number in one unit
cell, $a$ is the s-wave scattering length and $\omega _{y}$ and $\omega _{z}$
are the trapping frequencies in the transverse directions. We consider a 1D
BEC confined in an elongated cigar-shaped trap with high transverse trapping
frequencies ($y$ and $z$ directions), while the trapping potential in the
longitudinal direction ($x$) is negligible. We also assume the interaction
coefficients between atoms are the same for different hyperfine states,
which is a very good approximation because their difference is very small
\cite{Nonlinear}.

Even in the presence of nonlinear terms, the solution of the GPE is still
the Bloch wave in a periodic optical lattice \cite{BiaoWu1,BiaoWu2}. The
repulsive interaction shifts the Bloch spectrum upwards, and modifies each
band dispersion and energy gap at the same time. However, isolated flat
bands still exist in the presence of nonlinearity, as shown in Fig. \ref%
{flatness}b where the flatness of the lowest band is plotted for $c=0.05$.
Compared with the linear case $c=0$, the flatness of the nonlinear flat band
decreases with increasing nonlinearity (Fig. \ref{flatness}d) and the
maximum flatness is shifted towards a smaller value of $\Omega $. A typical
example of the nonlinear Bloch spectrum with an isolated flat band is shown
in the inset of Fig. \ref{flatness}b.

\begin{figure}[t]
\includegraphics[width=1\linewidth]{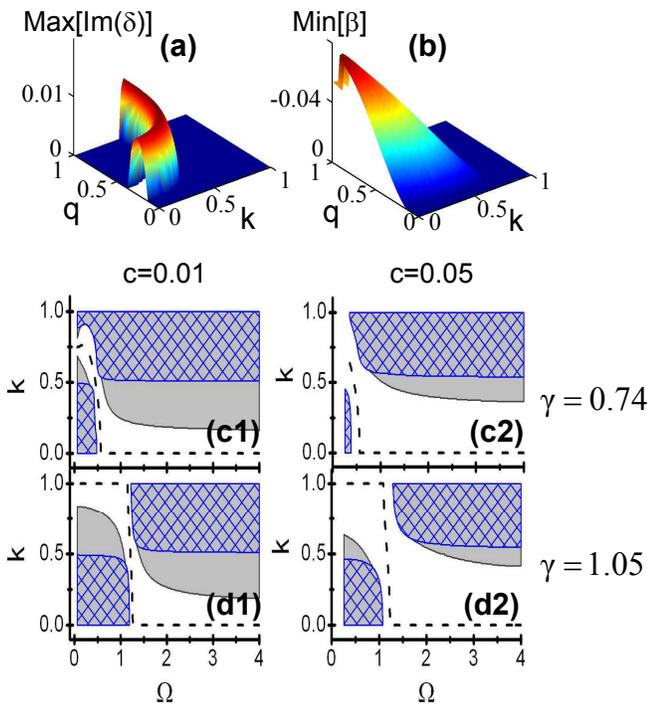}
\caption{(Color Online) Instability of the lowest Bloch band. (a) and (b):
Plots of the maximum of Im($\protect\delta $) for dynamical instability (a)
and the negative maximum of $\protect\beta $ for Landau instability (b). $%
V_{0}=1$, $c=0.01$, $\protect\gamma =1.05$, and $\Omega =0.8$. (c1), (c2),
(d1) and (d2): instability regions in the ($k$, $\Omega $) plane. Dashed
lines: the minimum of the lowest band. Blue grids: dynamical instability
regions. Gray shadows: Landau instability regions.}
\label{instability}
\end{figure}

The combination of nonlinear interaction and flat bands may lead to various
important phenomena, one of which is the superfluidity of the BEC in SO
coupled optical lattices. For BEC in optical lattices, the breakdown of
superfluidity may be caused by two different types of instabilities of the
BEC, dynamical instability and Landau instability, both of which have been
extensively studied in theory and experiments \cite%
{BiaoWu1,BiaoWu2,Smerzi,Machholm,Burger,Fallani,Morsch,Hooley,Dutton,Matt,Modugno}%
. The existence of SO coupling and flat bands may significantly modify the
superfluidity of the BEC. The stability analysis can be performed through
Bogoliubov theory, where quasiparticle excitations induced by perturbations
are taken into account through a small modification of the wavefunction $%
\Psi _{i}(x,t)=[\Phi _{i}(x)+\Delta \Phi _{i}(x,t)]\exp (-i\mu t+ikx)$,
where $\Phi _{i}(x)$ is the ground state of the BEC, $\Delta \Phi
_{i}(x,t)=u_{i}(x)\exp (iqx-i\delta t)+w_{i}^{\ast }(x)\exp (-iqx+i\delta
^{\ast }t)$, $q$ and $\delta $ are the wavevector and energy of the
quasiparticle excitations, while $u_{i}$ and $w_{i}$ are the quasiparticle
amplitudes. Substituting the modified wavefunction into the GPE, and
linearizing the GPE with respect to $u_{i}$ and $w_{i}$, we obtain
Bogoliubov-de Gennes (BdG) equations $\delta \varphi =\mathcal{M}\varphi $
with $\varphi =(u_{1},w_{1},u_{2},w_{2})^{T}$, where the matrix%
\begin{equation}
\mathcal{M}=%
\begin{pmatrix}
\mathcal{A}_{12} & \mathcal{B}_{12} \\
\mathcal{B}_{21} & \mathcal{A}_{21}%
\end{pmatrix}%
,
\end{equation}%
with%
\begin{eqnarray*}
\mathcal{A}_{mn} &=&%
\begin{pmatrix}
\mathcal{L}^{(mn)}(q+k) & c\Phi _{m} \\
-c\Phi _{m}^{\ast } & \mathcal{L}^{(mn)}(q-k)%
\end{pmatrix}%
, \\
\mathcal{B}_{mn} &=&%
\begin{pmatrix}
\Omega +c\Phi _{m}\Phi _{n}^{\ast } & c\Phi _{m}\Phi _{n} \\
-c\Phi _{m}^{\ast }\Phi _{n}^{\ast } & -\Omega -c\Phi _{m}^{\ast }\Phi _{n}%
\end{pmatrix}%
.
\end{eqnarray*}%
Here $\mathcal{L}^{(mn)}(k)=-1/2(\partial /\partial x+ik)^{2}+V_{0}\sin
^{2}(x)-i\gamma (\partial /\partial x+ik)-\mu +2c|\Phi _{m}|^{2}+c|\Phi
_{n}|^{2}$. Because the matrix $\mathcal{M}$ is not Hermitian, its
eigenvalues may be imaginary. The dynamical instability is defined if $%
\mathcal{M}$ has one or more non-zero imaginary eigenvalues. In this case,
the instability is characterized by the exponential growth of the
perturbation. The nonlinear Bloch wave is dynamically stable if all
eigenvalues are real numbers. On the other hand, the Landau instability can
be studied by solving the BdG equation \cite{BiaoWu1}, $\beta \varphi =\tau
_{z}\mathcal{M}\varphi $ with $\tau _{z}=\mathbf{I}\otimes \sigma _{z}$. The
nonlinear Bloch wave $\Phi _{i}$ is said to be Landau instable if one or
more eigenvalues of $\tau _{z}\mathcal{M}$ are negative. Physically, the
nonlinear Bloch wave with Landau instability is not the local energy minimum
of the system.

We systematically study the stability of nonlinear Bloch waves at the lowest
Bloch band for various parameters. A typical example of the dynamical and
Landau instability of the lowest band is shown in Figs. \ref{instability}a
and \ref{instability}b. Due to the symmetry in the plane ($q,k$), we only
show the region $0\leq q\leq 1$ and $0\leq k\leq 1$. In Fig. \ref%
{instability}a, the maximum of the imaginary part of $\delta $ is shown. The
BEC in the region with non-zero values is dynamically unstable. There is a
critical $k_{c1}$ beyond which Bloch waves become dynamically stable. In
Fig. \ref{instability}b, the negative maximum of $\beta $ is plotted, and
non-zero values indicate the Landau instability. There is also a critical $%
k_{c2}$ beyond which Bloch waves are the local energy minimum.

In Figs. \ref{instability}c and \ref{instability}d, we plot the dynamical
and Landau instability region for different nonlinearity, SO coupling and
Zeeman field. For BEC in a regular optical lattice, the energy minimum of
the lowest band locates at $k=0$, and the Bloch waves in the region around $%
k=0$ are stable. While in SO coupled optical lattices, the energy minimum of
the lowest band may not locate at $k=0$, therefore we expect the stability
domains should change accordingly, as clearly shown in Figs. \ref%
{instability}c and \ref{instability}d. For $\gamma =0.74<1$ in Figs. \ref%
{instability}c1 and \ref{instability}c2, the minimum of the lowest band
(dashed lines) shrinks to $k=0$ with increasing $\Omega $, and the abrupt
change of the energy minimum corresponds to the flat band region. When $%
\gamma $ is close to one half of the first BZ, e.g., $\gamma =1.05$ in Figs. %
\ref{instability}d1 and \ref{instability}d2, the energy minimum initially
stays at the edge of the first BZ $k=1$. In the flat band region, the energy
minimum quickly moves to $k=0$. For a larger $\gamma $, the energy minimum
initially increases from a value smaller than $k=1$ to the edge with
increasing $\Omega $, stays there for certain range of $\Omega $, and then
suddenly moves to $k=0$. The numerical results agree with the natural
expectation that Bloch waves surrounding the minimum of the lowest band are
stable, as shown in Fig. \ref{instability}c and \ref{instability}d. However,
we see the whole band is stable in the flat band region, which means that
the superfluidity of BEC with any momentum in the flat band is conserved.
There are another two properties: 1) the region of dynamic instability is
always smaller than the region of Landau instability; 2) the stable region
increases for a larger nonlinear coefficient $c$. These two properties are
the same as those for BEC in regular lattices \cite{BiaoWu1,BiaoWu2}.

\section{Conclusion}

In summary, we show that the combination of SO coupling, Zeeman field and
optical lattice can generate flat ground state energy bands where the
superfluid of the BEC is stable in the whole band region. Our proposed SO
coupling mechanism, when generalized to 2D, may provide an experimentally
feasible route for generating chiral flat bands and studying relevant
fractional quantum Hall insulator physics. The stable superfluidity in the
whole ground state band may lead to other interesting phenomena that have
not been explored in regular optical lattices, such as dissipationless Bloch
oscillation of BEC.

\textbf{Acknowledgement:} Y.Z. is supported by ARO (W911NF-09-1-0248), and
AFOSR (FA9550-11-1-0313). C.Z. is supported by DARPA-YFA (N66001-10-1-4025),
ARO (W911NF-12-1-0334), and NSF-PHY (1104546).



\begin{thebibliography}{99}
\bibitem{Nonlinear} The scattering lengthes for different hyperfine spin
states are $a_{11}=a_{12}=100.4a_{B}$ and $a_{22}=100.86a_{B}$ \cite{Lin}
with $a_{B}$ the Bohr radius. We use typical experimental parameters $\omega
_{y}=\omega _{z}=120Hz$, and the atom number in one unit cell $N=35$ which
corresponds approximately to $6\times 10^{4}$ total atom number. The
corresponding scaless parameters $c_{11}=c_{12}=0.0506$, $c_{22}=0.0508$ and
their difference is negiligible.
\end{thebibliography}
\end{document}